\documentclass[aps,pre, twocolumn, groupedaddress]{revtex4-1}
\usepackage{lipsum}
\usepackage{mathtools}
\usepackage{graphicx}
\usepackage{dcolumn}
\usepackage{amsmath}    
\usepackage{amssymb}
\usepackage{bm}
\usepackage{hyperref}
\usepackage{latexsym}
\usepackage{verbatim}
\usepackage{color}
\usepackage{float}
\usepackage[caption=false]{subfig}
\usepackage{epstopdf}

\def\beq{\begin{equation}}
\def\eeq{\end{equation}}

\def\beq{\begin{equation}}                          
\def\eeq{\end{equation}}                          
\def\bea{\begin{eqnarray}}                          
\def\eea{\end{eqnarray}}

\DeclareRobustCommand{\uvec}[1]{{%
  \ifcsname uvec#1\endcsname
     \csname uvec#1\endcsname
   \else
    \bm{\hat{\mathbf{#1}}}%
   \fi
}}

\preprint{}

\begin{document}

\title{Active nematic gel with quenched disorder}

\author{Sameer Kumar}
\email[]{sameerk.rs.phy16@itbhu.ac.in}
\affiliation{Department of Physics, Indian Institute of Technology (BHU), Varanasi, U.P. India - 221005}

\author{Shradha Mishra}
\email[]{smishra.phy@itbhu.ac.in}
\affiliation{Department of Physics, Indian Institute of Technology (BHU), Varanasi, U.P. India - 221005}
\date{\today}

\begin{abstract}

With quenched disorder, we introduce two-dimensional active nematics suspended in an incompressible fluid. We write the coarse-grained hydrodynamic equations of motion for slow variables, viz. density, orientation and flow fields. The quenched disorder is introduced such that it interacts with the local orientation at every point with some strength. Disorder strength is tuned from zero to large values. We numerically study the defect dynamics and system's kinetics and find that the finite disorder slows the ordering. The presence of fluid induces large fluctuation in the orientation field, further disturbing the ordering. The large fluctuation in the orientation field due to the fluid is so dominant that it reduces the effect of the quenched disorder. We have also found that the disorder's effect is almost the same for both the contractile and extensile nature of active stresses in the system. This study can help {\color{black} to} understand the impact of quenched {\color{black} disorder} on the ordering kinetics of active gels with nematic interaction among the constituent objects.   
\end{abstract}
\maketitle

\section{ Introduction}

In birds flock, fish's school, cytoskeletal filaments, migrating cells, etc.,  complex pattern formation, coherent motion and the spatiotemporal changes are fascinating visual events. Such interesting phenomena originate from the active nature of the systems composed of living particles. In an active system, each particle is driven by an active force and drives the system away from the equilibrium \cite{ProstNatPhys2015}. The active {\color{black} systems include} micro-organisms like cytoskeletal filaments \cite{doostmohammadi2016}, bacteria colonies\cite{CopenhagenNatPhys2021}, cells in tissue \cite{FriedlNatPhys2009} and macro-organisms like fish schools \cite{VicsekPRL1995}, bird flocks \cite{TonerPRL1995}, etc. Theoretical and experimental studies of active systems revealed many emergent behaviors, such as large density fluctuations \cite{Narayan2007, Chate2010}, spontaneously flocking states \cite{AditiPRL2002, KrusePRL2004, VoituriezEPL2005, VoituriezEPL2006, GiomiPRL2008}, strange rheological and structural properties \cite{CatesPRL2008, GiomiPRE2010} and spatiotemporal patterns that are not seen in passive complex fluids \cite{ChatePRL2006, ShradhaPRL2006}. The flocking phase of elongated active objects originates from their mutual alignment based on their head and tail symmetry. Polar objects like birds, fishes, etc., can order in polar or nematic fashion \cite{TonerTu1995, TonerTu1998, ZhangPNAS2018}, whereas apolar objects like melanocyte cells, m. xanthus bacteria form only nematic order (active nematics) \cite{KemkemerEPJE2000, DoostmohammadiNatComm2016}. Unlike polar objects, apolar objects do not distinguish between head and tail, i.e., the alignment unit vector ${\bm \nu}$ is invariant under the transformation  ${\bm \nu} \rightarrow -{\bm \nu}$ \cite{ShradhaPRL2006, MarchettiRMP2013}. 


\textcolor{black} {The Physics of active nematics gel} has gained a significant attention in the recent years, where the growth properties and defects dynamics are studied under various conditions, e.g., effect of underlying friction of the substrate and the  turbulence in the background fluid \cite{ProstNatPhys2015, SanthosJStatPhys2020, ThampiPRE2014, ZhangPNAS2018, ZhangPNAS2021, ThijssenPNAS2021, GuillamatPNAS2016, MartinezNatPhys2019}.

Inhomogeneity or disorder can play a crucial {\color{black} role} in the ordering of active systems; they can reduce the ordering as well as enhance the system dynamics according to their nature  \cite{ChepizhkoPRL2013, ReichhardtNatPhys2016, MorinNatPhys2017, RakeshDasPRE2018, DasEPL2018, PeruaniPRL2018, SinghJStat2021, SameerPRE2021}. In \cite{ChepizhkoPRL2013, ReichhardtNatPhys2016, MorinNatPhys2017}, authors have found that the presence of obstacles reduces the ordering and also breaks the ordered phase if their density is high.  These studies mainly address the impact of inhomogeneities on the ordering of polar particles. Still, quite a few studies address the effect of disorder in apolar active particles, \textcolor{black}{e.g., our previous work on dry active nematics with quenched disorder \cite{SameerPRE2020}.} 
\textcolor{black}{In \cite{ZhangPNAS2018, GuillamatPNAS2016, ThijssenPNAS2021}, authors experimentally studied the active nematic gel with some inhomogeneity, e.g., rigid microtubules (MT) in the the suspension of active filaments \cite{ZhangPNAS2018},  MTs based active nematic suspension in the presence of external magnetic field \cite{GuillamatPNAS2016} and active nematics flow in the presence of submerged microstructures \cite{ThijssenPNAS2021}, etc. Motivated form these works, in this study, we have investigated the impact of quenched disorder in wet-active nematics (or active nematic gel).} 
\textcolor{black}{Previous studies that address the defects kinetics in active nematic gel  \cite{HemingwaySoftMat2016, ThampiRSPTA2014, GiomiPRL2013, GiomiPRL2011, DoostmohammadiNatComm2016} are done mainly for clean systems, therefore, this study provides a thorough understanding of effect of inhomogeneity in an active nematic gel.}
\\

We use hydrodynamic equations of motion based on the continuum model \cite{TonerPRL1995, TonerTu1998} to study the two-dimensional  active nematics suspended in an incompressible fluid \cite{GiomiPRL2011, GiomiPRL2013,  MarchettiRMP2013, DoostmohammadiNatComm2016}, with quenched inhomogeneity in the  orientation field \cite{SameerPRE2020}. The equations are written in a coarse-grained description for the density field $\rho({\bf r},t)$, orientation field or nematic order parameter $\mathcal{Q}({\bf r},t)$, and the velocity of the flow field  ${\bf v}({\bf r},t)$. A coarse-grained study of active nematic gel in the presence of quenched disorder, ${\bf h}$, shows that the disorder slows the ordering kinetics in the system. The presence of fluid induces large fluctuations in the orientation field that reduces the effect of quenched disorder; still, large fluctuations in the orientation field due to fluid {\color{black} are sufficient enough to} delay the defects annihilation, which results in the slow ordering kinetic. This study can help in understanding the effect of quenched disorder and the flow field  strength in the naturally relevant systems, such as cytoskeletal suspensions in an incompressible fluid in the presence of unavoidable quenched inhomogeneity. \\

We divide the rest of the article in the following manner. In Sec. \ref{model}, we discuss the model and the numerical details; in Sec. \ref{results} we discuss the results and finally summarize in Sec. \ref{summery}.

\section{Model and Numerical details}
\label{model}
We write the hydrodynamic equations of motion for active nematics with quenched disorder suspended in an incompressible fluid referred  as ``active nematic gel'' in two dimensions. 
These equations are formulated in terms of local density field $\rho({\bf r},t)$, velocity of the flow field  ${\bf v(r},t)$, and the nematic order parameter $\mathcal{Q}_{ij}=S(\nu_i \nu_j-\frac{1}{2}\delta_{ij})$, where  ${\bf \nu}$ is the unit director and $i=1,2$ in two dimensions. $\mathcal{Q}_{ij}({\bf r},t)$ is uniaxial traceless and symmetric and hence have only two independent components in two dimensions. The disorder in the system is added to the $\mathcal{ Q}$ equation only. For simplicity we  write the hydrodynamic equation of motion for incompressible fluid, i.e., with ${\bf \nabla \cdot v}=0$. The density equation is given as, 

\begin{equation}
\centerline{$\frac{D \rho}{Dt} =\partial_i[\mathcal{D}_{ij}\partial_j \rho+\alpha_1 \rho^2 \partial_j \mathcal{Q}_{ij}],$}
\label{eq: 1}
\end{equation}
where, $\frac{D}{Dt}=[\partial_{t} +{\bf v} \cdot {\bf \nabla}]$ indicates the material derivative, $\mathcal{D}_{ij}=(D_0 \delta_{ij}+D_1 \mathcal{Q}_{ij})$ is the anisotropic diffusion coefficient term with constant $D_0$ and $D_1$. Equation for the flow field  is,
\begin{equation}
\centerline{$ \frac{D v_i}{Dt} =\eta \partial_i^2v_i-\partial_i p +\partial_j \sigma_{ij},$}
\label{eq: 2}
\end{equation}
where, $\eta$ is viscosity, $p$ is the pressure and $\sigma_{ij}$ is the stress tensor. We keep the density of the fluid equal to one.  Finally,  the equation for the orientation field or the nematic order parameter field is given as, 

\begin{equation}
\centerline{$\frac{D \mathcal{Q}_{ij}}{Dt}=\lambda S u_{ij}+\mathcal{Q}_{ik}\omega_{kj}-\omega_{ik}\mathcal{Q}_{kj}+\gamma^{-1}\mathcal{H}_{ij}+H_{ij}^{'}$} 
\label{eq: 3}
\end{equation}

  where, $\lambda$ is the flow  field parameter similar to what is used in \cite{WuARMA2019, DoostmohammadiNatComm2016, GiomiPRL2013} (larger the value of $\lambda$ stronger will be the effect of the fluid) , $u_{ij}=\frac{1}{2}(\partial_i v_j+\partial_j v_i)$ and $\omega_{ij}=\frac{1}{2}(\partial_i v_j-\partial_j v_i)$     are the symmetrized rate of strain tensor and vorticity, respectively. The molecular field $\mathcal{H}_{ij}$ embodies the relaxational dynamics of the nematic phase (with $\gamma$ as the rotational viscosity) and can be obtained from the variation of the Landau-de Gennes free energy of a two-dimensional nematic, $\mathcal{H}_{ij}=-\frac{\delta \mathcal{F} }{\delta Q_{ij}}$, with 

\begin{equation}
\centerline{$\frac{\mathcal{F} }{K}=\int dA[\frac{1}{4}(\rho-\rho_c)tr{\mathcal{ Q}}^2 +\frac{1}{4}\rho (tr{\mathcal{ Q}}^2)^2 +\frac{1}{2} \vert \nabla {\mathcal{ Q}} \vert^2]$}
\end{equation}

where, $K$ is an elastic constant with dimension of energy, $\rho_c$ is the critical density for isotropic-nematic transition, so that in ordered steady state scalar order parameter,  $S=\sqrt{1-\frac{\rho_c}{\rho}}$.  The quenched disorder is introduced  as {\em random field } in the free energy density $\mathcal{F} = -{\bf \mathcal{Q}} : ({\bf h}{\bf h}-\frac{{\bf I}}{2})$. We define quenched disorder as,  $ H_{ij}^{'}  = (h_i h_j-h_0^2\frac{1}{2}\delta_{ij})$, where,  $h_{i}=h_0(cos\phi,sin\phi)$ with $h_{0}$ as the disorder strength and $\phi({\bf r})$ is a uniform random angle between $(0, 2\pi)$ with mean zero, quenched in time and space correlation  $\langle \phi({\bf{r}}) \phi({\bf{r'}})\rangle= \delta ({\bf{r}}-{\bf{r'}})$.

Finally, the stress tensor in eq. (\ref{eq: 2}) $\sigma_{ij}=\sigma_{ij}^{r}+\sigma_{ij}^{a}$ is the sum of  elastic stress due to nematic elasticity, $\sigma_{ij}^{r}=-\lambda S \mathcal{H}_{ij} +\mathcal{Q}_{ik}\mathcal{H}_{kj}-\mathcal{H}_{ik}\mathcal{Q}_{kj}$, and $\sigma_{ij}^{a}=\alpha_2 \rho^2 \mathcal{Q}_{ij}$ is the active stress. 
 Activity yields a curvature induced active current ${\bf j}_a=\alpha_1 \rho^2 {\bf \nabla \cdot \mathcal{Q}}$ in equation (\ref{eq: 1}). The $\rho^2$ dependence of the active stress and current is appropriate for systems where activity arises from pair interactions among the filaments via cross-linking motor proteins \cite{SanchezNature2012}. The sign of $\alpha_2$ depends on whether the active particles generate contractile ($\alpha_2 > 0$) or extensile ($\alpha_2 < 0$) stresses, while we always keep $\alpha_1 > 0$.

  Eqs. (\ref{eq: 1} -  \ref{eq: 3}) written in dimensionless units by rescaling all lengths by the length  of the 
particle and time  by the collision time and are of the same form as derived
from the microscopic rule-based model in \cite{BertinNJoP2013}, with fluid flow \cite{GiomiPRL2011, GiomiPRL2013}  and an additional term due to 
{\em quenched disorder} as in the dry case \cite{SameerPRE2020}.

The random field introduced in our current model is similar to the random field
 in XY-model (RFXY-model) \cite{imryma1975}. 
Hereafter we refer our model as random field wet active nematics or RFWAN when $h_0\ne 0$, and  clean-wet active nematics (clean-WAN) for $h_0=0$. \textcolor{black}{We keep the activity moderate so that the system does not get into the high turbulence regime,  and  the effect of  quenched disorder remains relevant}.

To perform the numerical integration of Eqs. (\ref{eq: 1} - \ref{eq: 3} ) we construct a  two-dimensional $L \times L$ square lattice 
with periodic boundary condition (PBC) and discretise the space and time derivatives   
using \textcolor{black}{explicit} Euler scheme ($\Delta x = 1.0$ and $\Delta t = 0.1$). Initially, we start with random 
homogeneous  density with mean, $\rho_0=0.75 > \rho_c$, random  orientation and homogeneous flow field. \\

Parameters in Eqs. (\ref{eq: 1}-\ref{eq: 3}) are $\alpha_1=0.2, 0.3, \ \vert \alpha_2 \vert=2 \alpha_1$, $D_0=1.0, \ D_1=0.5$, $ \gamma=1.0, \  \eta=2.0$. We study the ordering kinetics of RFWAN for different values of disorder strength $h_0 \in [0.0,0.2]$ and flow  field parameter $\lambda =  0.1, 10, 20$  and   $100$.   One simulation time is counted after update of Eqs. (\ref{eq: 1} - \ref{eq: 3}) for all
lattice points. Also, the data in section \ref{kinetics} are averaged over $15$ independent configurations of   $\phi$. We check the stability of the code for the chosen set of parameters by calculating the fluctuation in the velocity field $v({\bf r},t)$ from its mean value, i.e., $\Delta v$. In $\Delta v$ vs. $t$ plot, we observe that $\Delta v$ show small fluctuation as $t \rightarrow \infty$, \textcolor{black}{see fig. \ref{fig:S1.7} in the Appendix \ref{appendixC}}.  \\

\begin{figure}      
{\includegraphics[width=\linewidth]{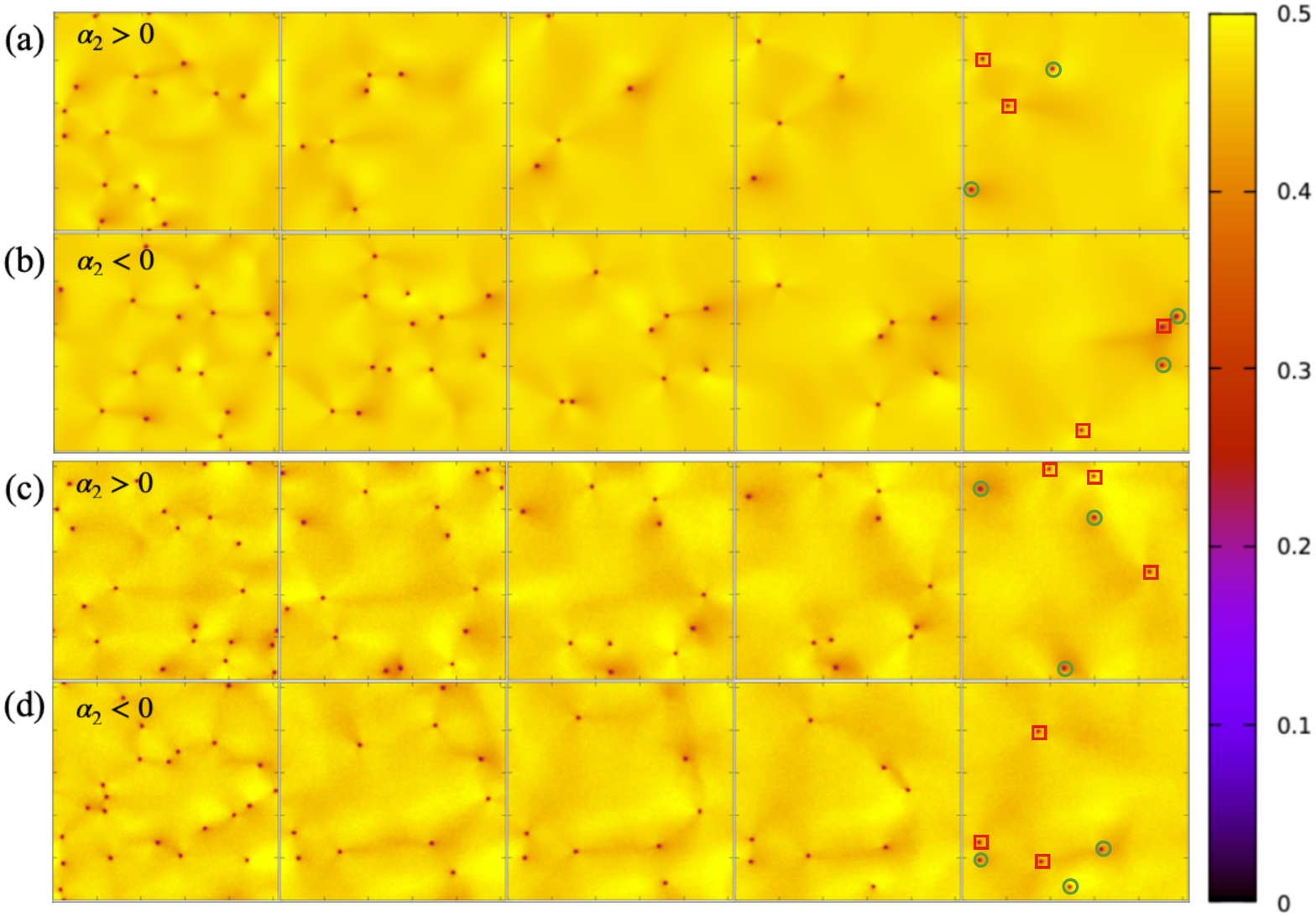}}  
\caption{ Snapshots of NOP at different simulation times for $\alpha_2>0$ (a,c) and $\alpha_2<0$ (b,d); for  disorder strengths $h_0=0.0$ (a,b) and $h_0=0.2$ (c,d). $+1/2$ defects are marked in red circles and $-1/2$ defects are marked in green squares in the rightmost snapshots. See the Supplementary material for the animation.}
\label{fig:102}
\end{figure}

\section{Results}
\label{results}
To characterise the system properties we calculate the magnitude of the nematic order parameter (NOP) defined as, $   \vert \mathcal{Q}_{ij} \vert =\frac{1}{2}\sqrt{ \mathcal{Q}_{11}^2  + \mathcal{Q}_{12}^2}$, where, $\mathcal{Q}_{11}=-\mathcal{Q}_{22}=\sum sin 2 \theta({\bf r},t)$ and $\mathcal{Q}_{12}=\mathcal{Q}_{21}=\sum cos 2 \theta({\bf r},t)$, here $\theta({\bf r},t)$ is the orientation field. We discuss the results in four subsections, first, we study the defect dynamics for different system parameters; second, we study the system kinetics; third, we study the effect of flow field and finally we study the the scaling properties of the system.
 \\
\subsection{ Defect dynamics} 
In two-dimensional active nematics, when the system is allowed to equilibrate, the ordering in the system takes place via the creation and annihilation of the topological defects of equal and opposite topological charges, i.e., $\pm 1/2$-defect pairs  \cite{DoostmohammadiNatComm2016}. $+1/2$ defects are asymmetric comet-like structures that act like motile particles and move convectively along the axis of asymmetry. In contrast, $-1/2$ defects have symmetric trefoil structures that only diffuse in the system. Further, the value of NOP is zero at the core of the defects; therefore, while approaching the ordered state, the defects pairs  get annihilated. \textcolor{black}{A brief comparison of wet active nematics with its passive counterpart is given in the Appendix \ref{appendixA}}.  In our previous study for the dry-RFAN, we find that a finite disorder in the system slows the dynamics of the $+1/2$ defects that result in slow coarsening \cite{SameerPRE2020}. We find the same observation in the presence of fluid also, where, with the finite disorder (RFWAN), we see more pair of defects (see fig. \ref{fig:102}(c,d)) than the clean case or clean-WAN (see fig. \ref{fig:102} (a,b)). Further, the effect of disorder is the same for both contractile (when $\alpha_2>0$) and exetensile (when $\alpha_2<0$) nature of active stresses, $\sigma^{a}$ (see eq. (\ref{eq: 2})). Fig \ref{fig:102}(a-d) show the snapshots of $\mathcal{Q}_{ij}$ at different simulation time. We can see the defects pairs and their annihilation as the simulation time increases. The defect annihilation is fast for $h_0=0.0$  (fig. \ref{fig:102}(a,b)) compared to case when $h_0=0.2$ (fig. \ref{fig:102}(c,d)), which suggest that in the presence of quenched disorder, we observe slow defect dynamics that can results in slow ordering. Further, we see almost the same number of defects pairs for both $\alpha_2>0$ and $\alpha_2<0$ without the disorder (fig. \ref{fig:102}(a,c)) and with disorder (fig. \ref{fig:102}(c,d)). We find the same observation in correlation length vs. time plot (discussed later). Therefore, we study the system's response only for $\alpha_2>0$ in the further results and discussion.\\

\begin{figure}      
{\includegraphics[width=\linewidth]{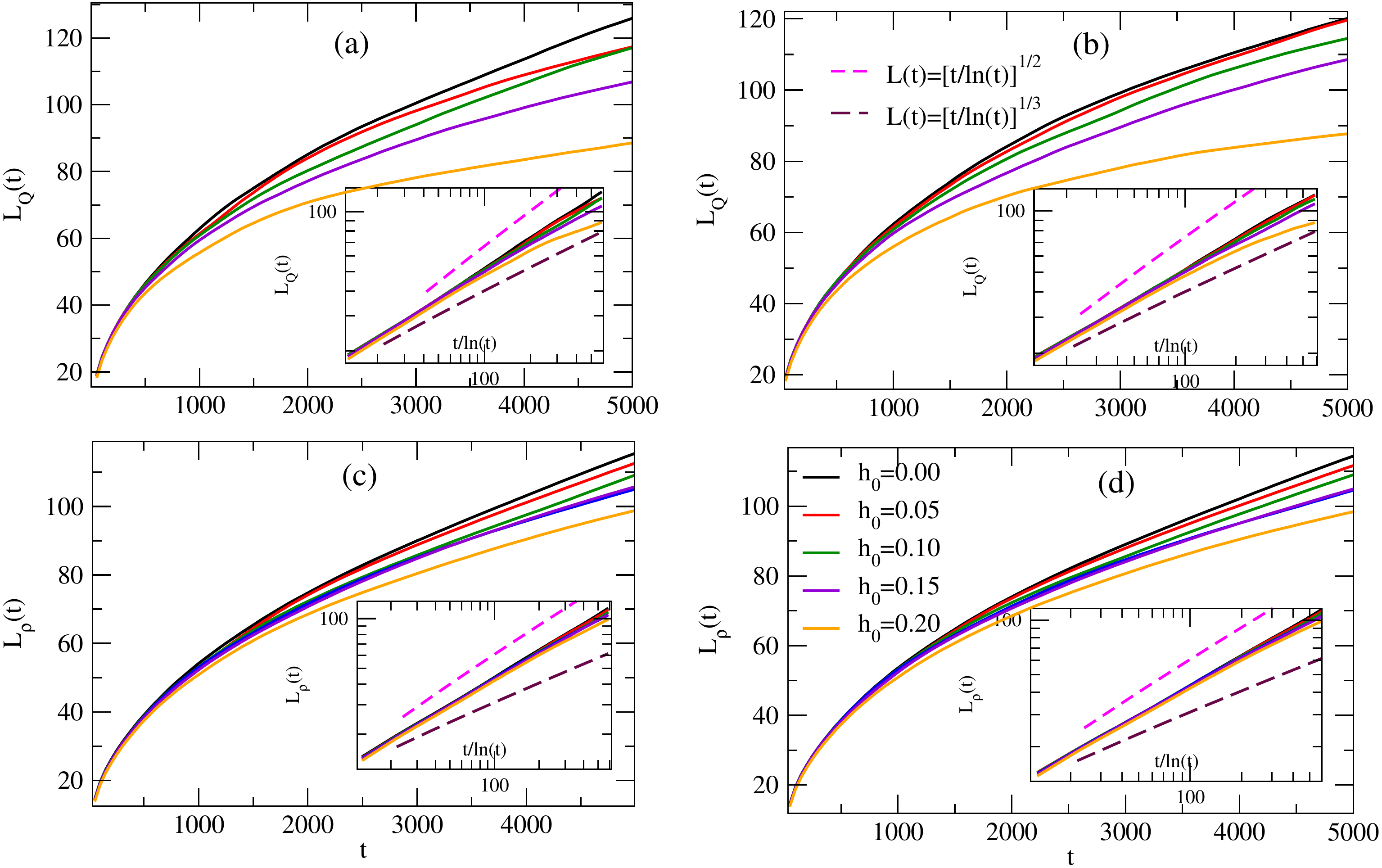}}   
\caption{Correlation length, $L_{\mathcal{Q}}$ vs. $t$ (a,b) and $L_{\rho}$ vs. $t$ ( c,d ) for different $h_0$. Plot in left panel (a,c)  are for contractile case with $\alpha_2=0.4$, whereas right panel (b,d) shows the plots for extensile case with  $\alpha_2=-0.4$. Also the value of flow field  parameter $\lambda=0.1$. Insets : Correlation length, $L_{\mathcal{Q},\rho}(t)$ vs. $t/\ln(t)$ on log-log scale. }
\label{fig:1}
\end{figure}

\begin{figure}      
{\includegraphics[width=\linewidth]{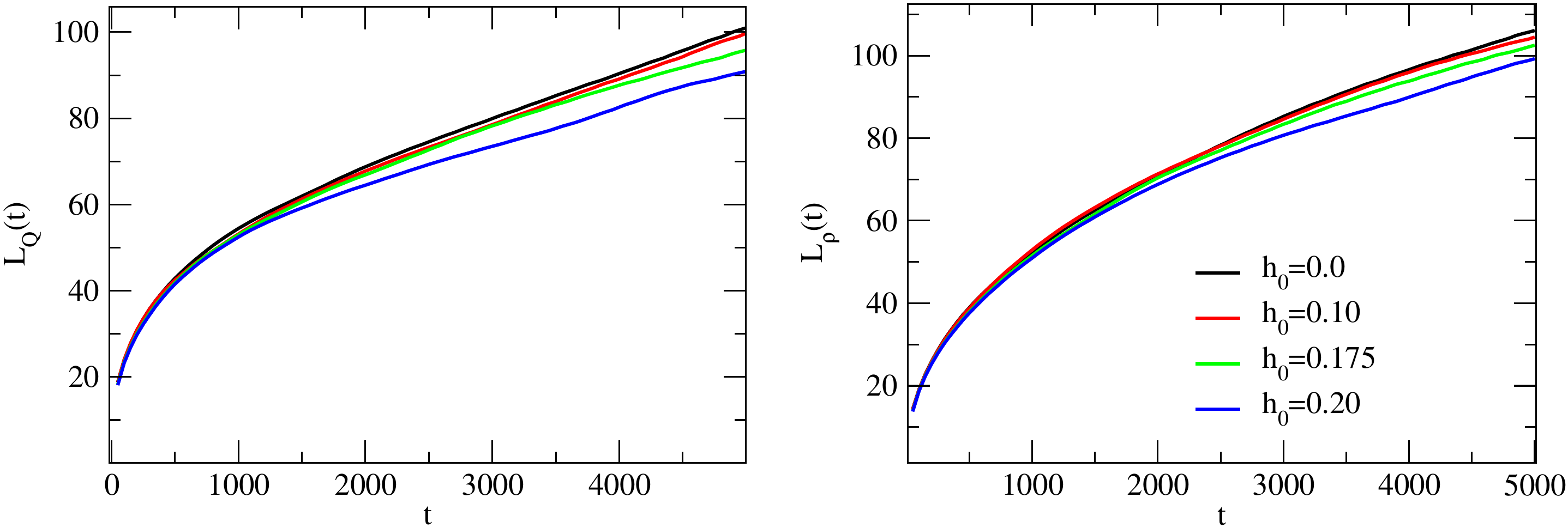}} 
\caption{ Correlation length $L_{Q}(t)$ vs. $t$ (left) and  $L_{\rho}(t)$ vs. $t$ for different strength of disorder $h_0$ and fixed flow field  parameter, $\lambda=100$. } 
\label{fig:20}
\end{figure}

\subsection{Kinetics} 
\label{kinetics}
To understand the effect of quenched disorder on the ordering kinetics, we calculate the correlation length for Order parameter field, $L_{\mathcal{Q}}(t)$ and density field, $L_{\rho}(t)$ and plot it for different strength of disorders, $h_0$. The correlation lengths ($L_{{\mathcal{Q}},\rho}(t)$) is defined as the length of the first zero crossing of the correlation function, $C_{{\mathcal{Q}},\rho}({\bf r},t)$ (see section \ref{corsening}). Fig. \ref{fig:1}(main), shows the plots of correlation lengths, $L_{\mathcal{Q}, \rho}(t)$ vs. time $t$ for different strengths of disorder, $h_0$. We see that as time increases, correlation length increases for a fixed strength of disorder. Further,  we observe that the correlation length (or the size of the ordered domain), for a fixed time, decreases as we increase the strength of the quenched disorder in the system. The impact of disorder is similar for both contractile ($\alpha_2>0$) and extensile ($\alpha_2<0$) cases, \textcolor{black}{which is robust for other value of activity, i.e., $\alpha_1=0.3$ (see fig. \ref{fig:S1.4} in the Appendix \ref{appendixB})}. This observation is different from what is observed for scalar active particles suspended in an incompressible fluid, where the growth of $L(t)$ is faster for extensile stress than that of contractile stress \cite{TiribocchiPRL2015}. Further, in fig. \ref{fig:1}(insets), we show the plot of  $L_{\mathcal{Q}, \rho}(t)$ vs. $t/\ln(t)$ (where $t/\ln(t)$ is the logarithmic correction \cite{BrayAdvPhys1994}) on $log-log$ scale. The correlation length grows as $L_{\mathcal{Q}, \rho}(t) \sim [t/\ln(t)]^{1/z}$ , where the dynamic growth exponent \cite{BrayAdvPhys1994} $z \simeq 2.0$ for $h_0=0.0$ and increases in range $(3.0>z>2.0)$ as we increase the value of $h_0$.    It again conveys that the quenched disorder in a two-dimension wet active nematics slows the ordering. {\textcolor{black}{In addition to above analysis, we did a brief comparison between wet active and passive nematics and find the order parameter field follows the same growth law for both passive and active cases, whereas in the passive nematics no growth is found for density field (see fig. \ref{fig:S1.3} in the Appendix \ref{appendixA})}}.
\\

Up to here, we have analyzed the effect of quenched disorder in RFWAN with a fixed flow field parameter ($\lambda=0.1$). Now, we explore the response of fluid and the effect of disorder for various strength of flow controlling parameters or the flow field parameter ($\lambda$) in the system.

\begin{figure}      
{\includegraphics[width=\linewidth]{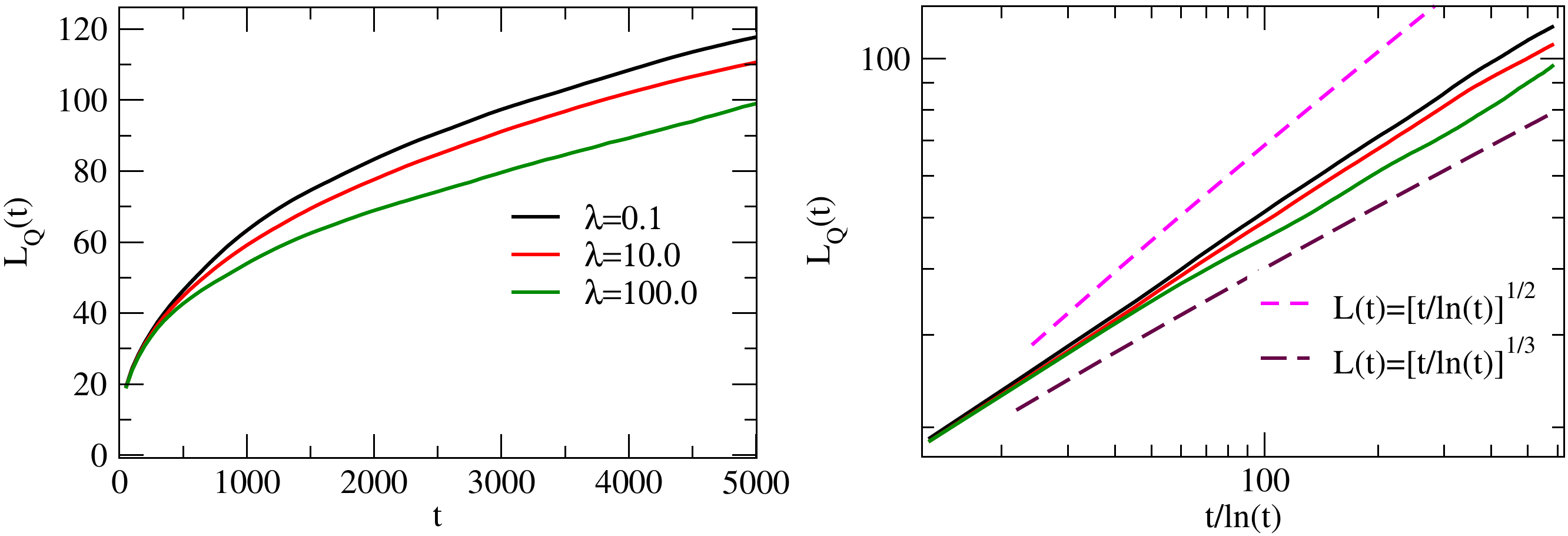}}   
\caption{Correlation length $L_{Q}(t)$ vs. $t$ on linear scale (left)  and log-log scale (right), for different $\lambda$ and fixed $h_0=0.1$ for $\alpha_2=0.4$.   }
\label{fig:21}
\end{figure}

\begin{figure}      
{\includegraphics[width=\linewidth]{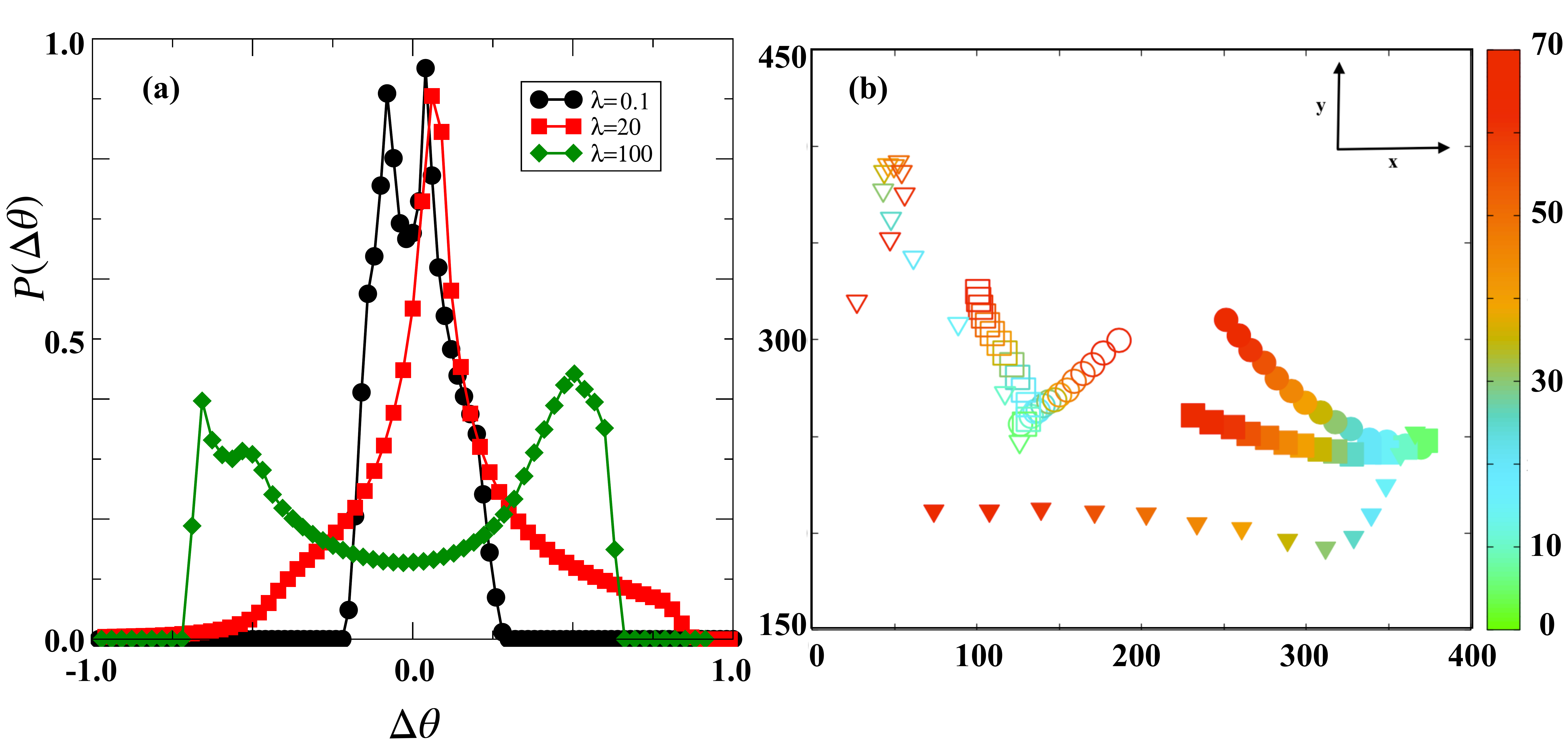}}   
\caption{(a) Probability distribution function $P(\Delta \theta)$ vs angle fluctuation $\Delta \theta$ for different values of $\lambda=0.1, 20.0 \ and \ 100.0$. Data on the x-axis is in units of $\pi/2$. We fix the strength of quenched disorder, $h_0=0.1$ and $\alpha_2=0.4$. Data is generated  when the system reaches the steady state and averaged over $10$ independent snapshots of  orientation field ($ \theta ({\bf r},t)$). (b) Trajectories of topological defects  for $\lambda=0.1$ (circles), $\lambda=20.0$ (squares) and  $\lambda=100.0$ (triangles). Filled data symbols are for $+1/2$ defects and open symbols represents $-1/2$ defects. Data is generated for $h_0=0.0$ and $\alpha_2=0.4$. Color bars shows the simulation time $t/1000$.}
\label{fig:22}
\end{figure} 

\subsection{Effect of fluid  in RFWAN}
\label{fluid effect}
To study the effect of fluid on the growth properties for different strengths of disorder in the system, we calculate the correlation length $L_{\mathcal{Q},\rho}(t)$ for different strengths of disorder $h_0$ and flow aligning parameter $\lambda=100.0$. The flow field  parameter $\lambda$ controls how the director field rotates in a shear flow and affects the flow and rheology of the systems. In fig. \ref{fig:20}, we show the plot of $L_{\mathcal{Q},\rho}(t)$ vs. $t$ for different strengths of quenched disorder ($h_0$). We again find that in the presence of disorder, correlation length (at time $t$) decreases as we increase $h_0$. \textcolor{black}{Still, the effects of the quenched disorder are reduced for $\lambda=100.0$  than the former case when $\lambda=0.1$ (see fig. \ref{fig:1}).}      Further, in fig. \ref{fig:21}, we plot the correlation length ($L_{\mathcal{Q}}(t)$) for a fixed disorder strength ($h_0=0.1$), and different values of $\lambda$. In this plot, we observe that the size of the ordered domain at a fixed time decreases as we increase $\lambda$. This effect can be understood by calculating the probability distribution function $P(\Delta \theta)$ for different values of flow field parameter ($\lambda$), where $\Delta \theta$ is  the angle fluctuation in the orientation field ($\theta$) from its mean ($\theta_0$). Fig. \ref{fig:22}(a) show the plot of $P(\Delta \theta)$ vs. $\Delta \theta$ for three different values of $\lambda=0.1, 20.0$ and $100.0$ and fixed $h_0=0.1$. From this plot, we observe that the fluctuation in the orientation field increases as we increase the value of the flow field   parameter. Also, for $\lambda=100.0$, we observe two distinct peaks that imply uncorrelated domains. Therefore,  we do not observe a homogeneous ordered phase in the steady-state for a large values of flow field   parameter.   This tells us that with an increase in the value of $\lambda$, the local orientation itself gets randomized, which causes large fluctuations in the nematic order parameter $\mathcal{Q}$, hence reducing the growth dynamics. This effect can be seen in the defect dynamics in fig. \ref{fig:22}(b), which shows the trajectories of $\pm 1/2$-defects for three different values of $\lambda=0.1, 20.0$ and $100.0$. We see that for $\lambda (=0.1$  and  $20.0)$, the trajectories of defects are smooth, whereas, for $\lambda=100.0$, it is distorted. Further, for  $\lambda=0.1$, the defects attract each other from the early time, but for higher values, i.e., $\lambda= 20.0$  and $ 100.0$, the defects initially move away from each other and later on come closer so that they can annihilate each. Therefore, for the large values of flow field  parameter, the defects annihilation becomes slow, which results in the slow ordering kinetics.   Therefore, since the quenched disorder is associated with the nematic order parameter field in the system, its effect is no more significant, and the local fluctuation dictates the dynamics in the nematic order parameter due to the fluid. Therefore,  we conclude that the quenched disorder negatively affects the growth kinetics in an active nematic gel. Still, the effect reduces as we increase the flow field  parameter in the system. \\   

\subsection{Correlation functions and scaling properties}
\label{corsening}

\begin{figure}      
{\includegraphics[width=\linewidth]{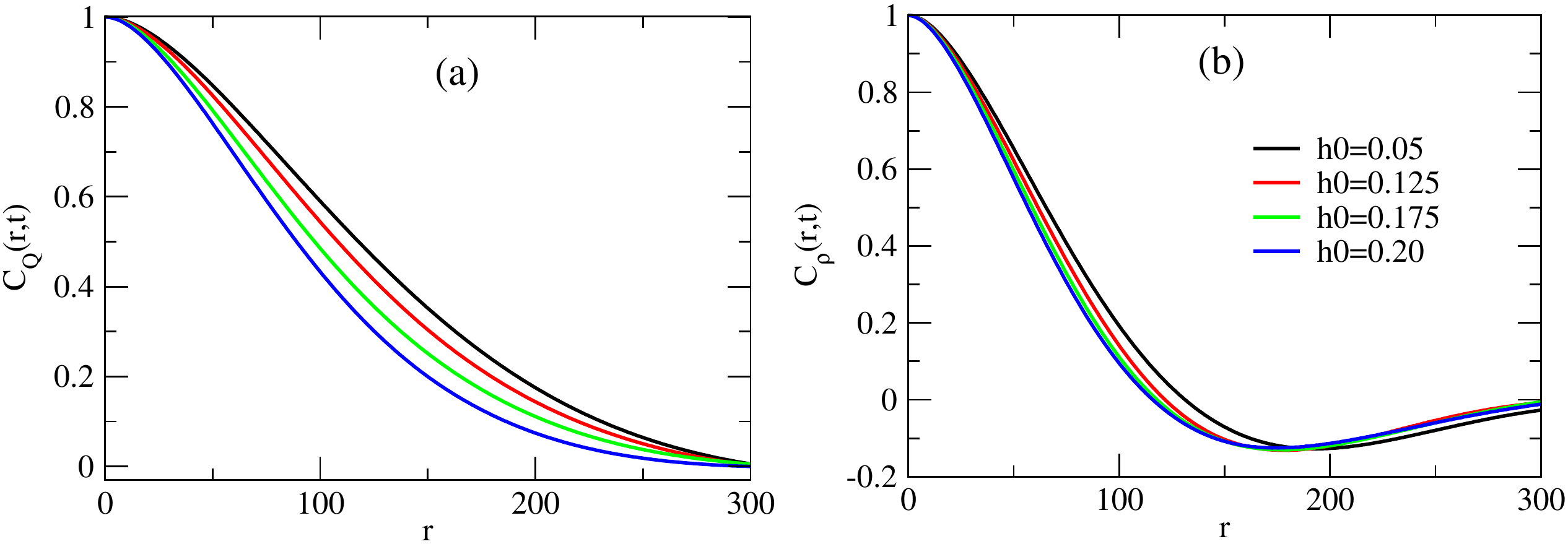}}  
\caption{(a) Two point orientation correlation function $C_{\mathcal{Q}}({\bf r},t)$ vs.  distance ${\bf r}(t)$  and  (b) density correlation function $C_{\rho}({\bf r},t)$ vs.  distance ${\bf r}(t)$ for different strength of quenched disorder ($h_0$). All data are taken for $\lambda=0.1$ at $t=5000$ and averaged over $15$ independent ensembles.} 
\label{supfig:3}
\end{figure}

\begin{figure}      
{\includegraphics[width=\linewidth]{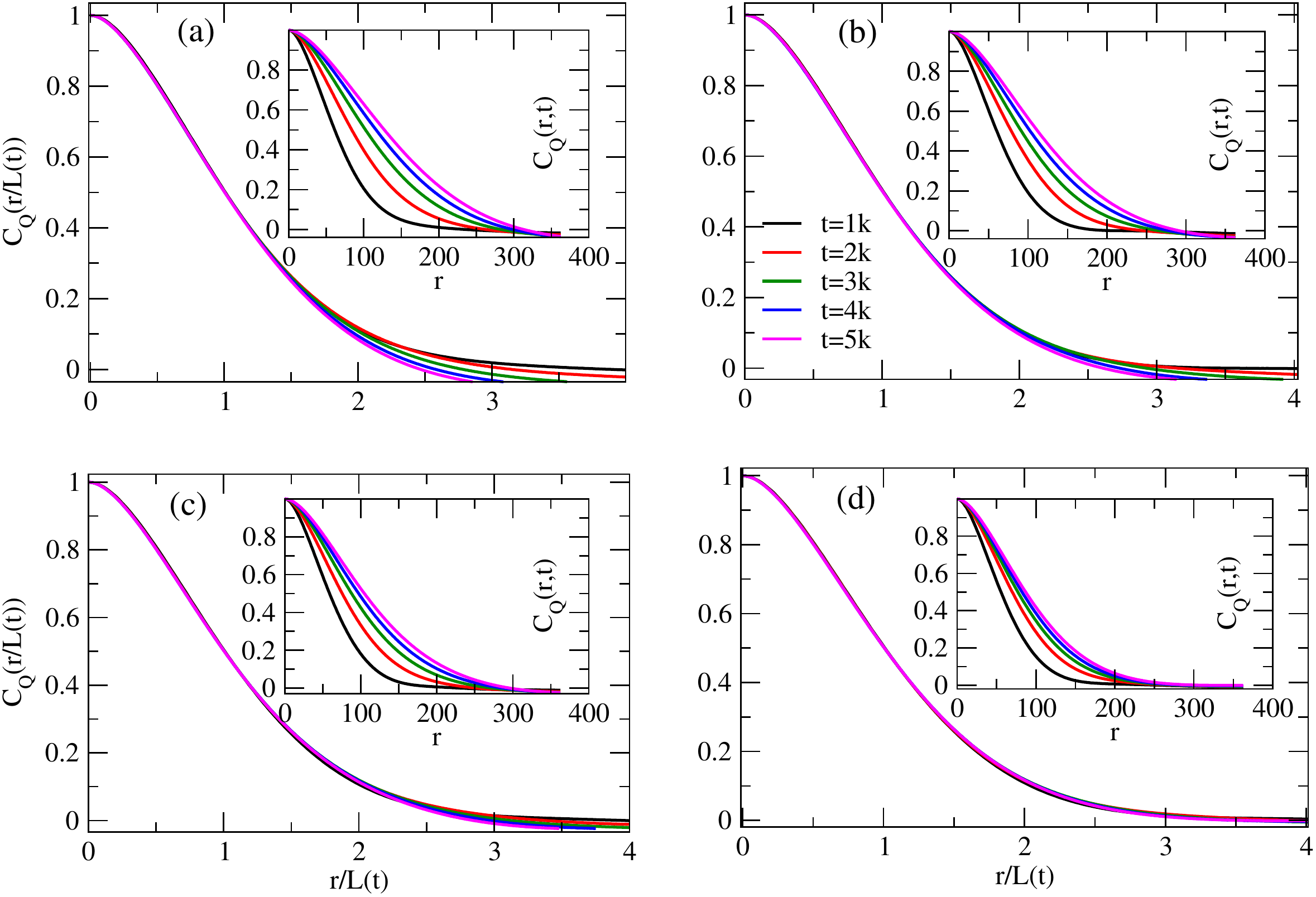}}
\caption{Scaled two point correlation function, $C_{\mathcal{Q}}({\bf r}/L_{\mathcal{Q}}(t),t)$ vs. scaled distance ${\bf r}/L_{\mathcal{Q}}(t)$  for  $h_0=0.0$ (a), $h_0=0.1$ (b), $h_0=0.15$ (c) and $h_0=0.2$ (d).  Insets: $C_{\mathcal{Q}}$ vs. $r$ at different simulation time. Data are taken for $\lambda=0.1$ and averaged over $15$ independent  realizations and $1k =1 \times 1000$.}
\label{supfig:4}
\end{figure}

We study the ordering kinetics and scaling properties of RFWAN for different disorder strengths ($h_0$).
We calculate the two point correlation functions  for orientation and density fields, $C_{\mathcal{Q},\rho}$, defined as,  $C_{\mathcal{Q}}({\bf{r}},t)=\langle\mathcal{ Q}({\bf{0}},t):\mathcal{Q}({\bf{r}},t)\rangle$ where $a:b=a_{ij}b_{ij}$ and, local density $\rho$, $C_{\rho}({\bf{r}},t)=\langle \delta \rho({\bf{0}},t) \delta \rho ({\bf{r}},t)\rangle$, where $\delta \rho({\bf{r}},t)=\rho({\bf{r}},t)-\rho_0$ is the deviation of the local density from the mean $\rho_0$. Fig. \ref{supfig:3} shows the plot for $C_{\mathcal{Q},\rho}({\bf r},t)$ vs. $r$ for different strengths of quenched disorder,$h_0$, and fixed at  simulation time $t$. From these plots we observe that the correlation among the particles decreases with distance. The decrease  in $C_{\mathcal{Q},\rho}({\bf r},t)$ is fast for large values of $h_0$ compare to small values. \textcolor{black}{The fast decay in correlation function with distance is due to the slow defect annihilation in the presence of quenched disorder.}

We again plot $C_{\mathcal{Q},\rho}({\bf r},t)$ vs. $r$ at different simulation time $t$ for  different values of  quenched disorder strengths $h_0$ in fig. \ref{supfig:4}(insets:a-d). We observe that the correlation in the orientation field $C_{\mathcal{Q}}({\bf r},t)$ and density field $C_{\rho}({\bf r},t)$ increases with time. Further, we scale the distance as $r \rightarrow r/L(t)$, where $L(t)$ is the correlation length, and plot $C_{\mathcal{Q}}({\bf r}/L_{\mathcal{Q}}(t),t)$ vs. scaled distance ${\bf r}/L_{\mathcal{Q}}(t)$ in fig. \ref{supfig:4}(main:a-d). We see that, for clean-WAN, the system shows dynamic scaling up to the distance equal to the correlation length and scaling not found for $r > L(t)$ (see fig. \ref{supfig:4}(a)). But, this behavior gradually disappears when we increase the strength of quenched disorder, and the system shows good dynamic scaling at larger distance $r > L(t)$ for $h_0=0.2$, see fig. \ref{supfig:4}(d).   \textcolor{black}{ To confirm this behaviour is not due to the system-spanning or the finite size effect, we check the scaling properties for early time, i.e., $t=2500$ , and we again find that the system shows dynamics scaling for $r>L(t)$ only in the presence of finite disorder \textcolor{black}{(see fig. \ref{fig:S1.5} in the Appendix \ref{appendixB})}. Also, the behaviour remains consistent for larger activity too, i.e., when $\alpha_1=0.3$ \textcolor{black}{(see fig. \ref{fig:S1.6} in the Appendix \ref{appendixB})}.} Therefore, these results suggest that, in RFWAN, scaling becomes better as we increase the strength of the quenched disorder $h_0$, which is surprising and different from the dynamic scaling properties observed in RFAN \cite{SameerPRE2020}, where the system shows good dynamic scaling for all the values of quenched disorder.

\section{Discussions} 
\label{summery}
We numerically studied the two-dimensional active nematics with quenched disorder suspended in an incompressible fluid. The quenched disorder is introduced in the orientation field, and we call it random field wet active nematics (RFWAN). Results from the numerical simulation suggest that in RFWAN, finite disorder slows down the defect annihilation, resulting in slow coarsening in the system. Effect of the quenched disorder is similar for both the contractile and extensile nature of the active stresses in the system.  
Further, the presence of fluid induces large fluctuations in the orientation field, due to which the defect annihilation slows. The disorder is introduced such that each particle feels quenched noise of fixed strength in its orientation; therefore, large fluctuations in the nematic order parameter due to the fluid reduces the disorder's effect.   We also find that the system shows dynamics scaling only  for large value of quenched disorder strength, which is a surprising result and can be a potential problem to explore.\\ 
This study reveals that, although the fluid in which apolar active particles are suspended reduces the quenched disorder's impact, it also disturbs the local ordering and consequently delays the coarsening. This work also encourages us to see the effect of fluid on the ordering of polar flocks in the presence of quenched disorder.\\

\section{Acknowledgement }
 S.M. and S.K. thank DST- SERB India, Grant No. ECR/2017/000659, for financial support. S.M. and S.K. thank Luca Giomi for his initial inputs in the problem. S.M. and S.K. thank Sanjay Puri for his useful comments.

\bibliographystyle{apsrev4-1}
\bibliography{references} 

\appendix

\begin{figure*}      
{\includegraphics[width=16 cm , height=10 cm]{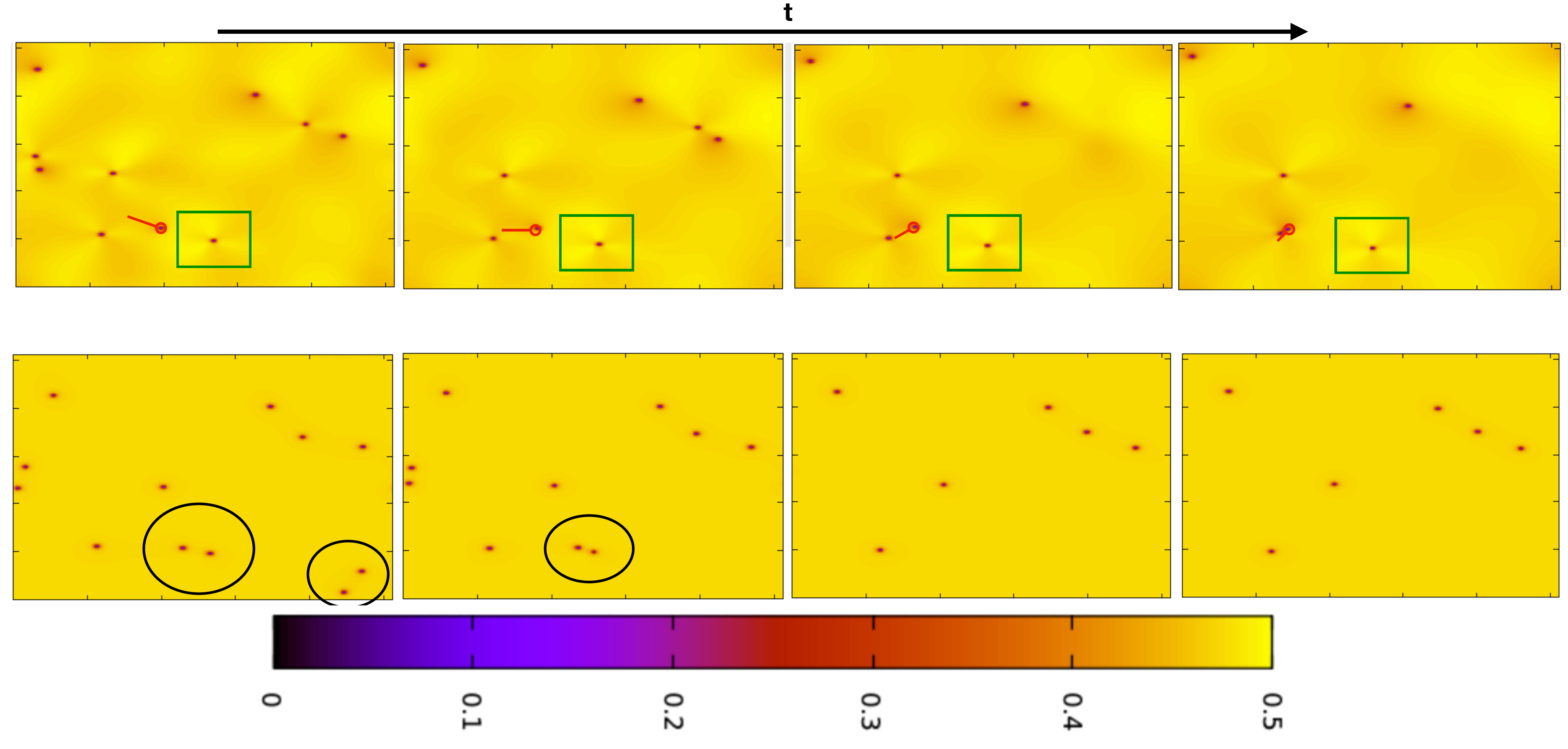}}   
\caption{\textcolor{black}{Figure shows the snapshots  at different time $t$ for nematic order parameter $\mathcal{Q}$ when $\alpha_1=0.2$ (top panel) and $\alpha_1=0.0$ (bottom panel). In the case when $\alpha_1$ is non-zero or the active case, defects have typical topological structure where $+1/2$ defect is asymmetric (red circle with stick) moves along the stick and the $-1/2$ defects have trefoil structure (inside green square) only diffuses. In contrast, when $\alpha_1=0.0$ or the passive case, there is no visible topological structure the $-1/2$ and $+1/2$ defects pairs (inside black circles), hence the direction on motion cannot be identified. All the snapshot are generated for clean case (i.e. $h_0=0.0.$). Axes of every snapshot is in range $[0,512]$. Animations can be seen in the supplementary multimedia files.}}
\label{fig:S1.1}
\end{figure*}

\begin{figure*}      
{\includegraphics[width=17 cm , height=8 cm]{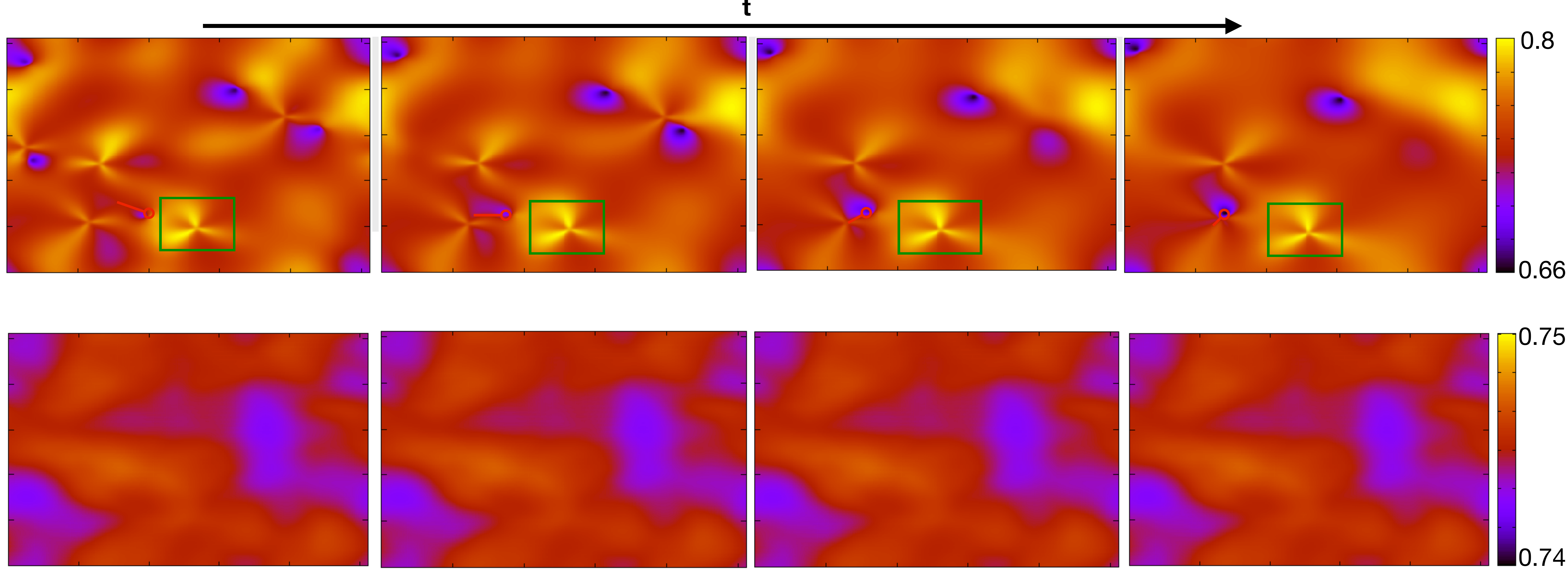}}   
\caption{\textcolor{black}{Figure shows the snapshots  at different time $t$ for the density field $\rho$ when $\alpha_1=0.2$ (top panel) and $\alpha_1=0.0$ (bottom panel). In the case when $\alpha_1$ is non-zero or the active case, defects have typical topological structure where $+1/2$ defect is asymmetric (red circle with stick) moves along the stick and the $-1/2$ defects have trefoil structure (inside green square) only diffuses. In contrast, when $\alpha_1=0.0$ or the passive case, there is no topological defects and the density field do not evolve with time. All the snapshot are generated for clean case (i.e. $h_0=0.0.$). Axes of every snapshot is in range $[0,512]$. Animations can be seen in the supplementary multimedia files.}}
\label{fig:S1.2}
\end{figure*}

\section{Active versus passive wet  nematics}
\label{appendixA}
\begin{figure}      
{\includegraphics[width=\linewidth]{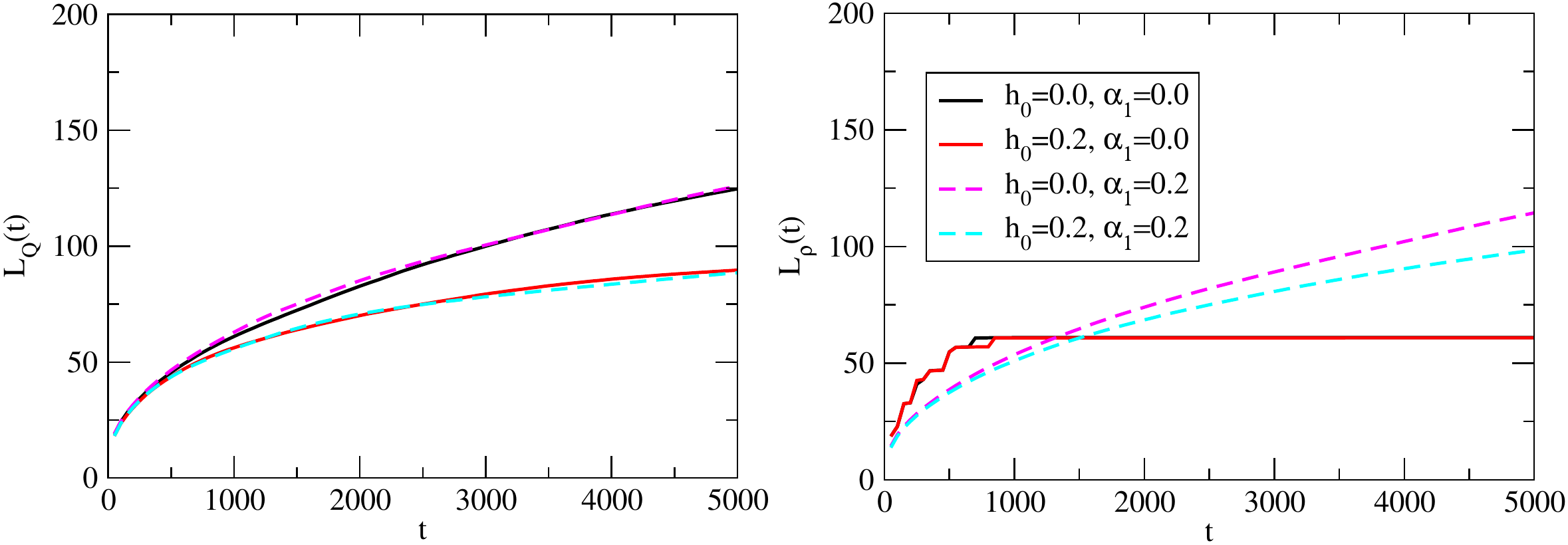}}   
\caption{Figure shows the correlation length plots, $L_{\mathcal{Q}}(t)$ (left) and $L_{\rho}(t)$ (right) vs. time for passive (when $\alpha_1=0.0$, solid lines) and active (when $\alpha_1=0.2$, dashed lines) cases for disorder strength $h_0=0.0$ and $h_0=0.2$.}
\label{fig:S1.3}
\end{figure}

 To make a comparison between the active and passive cases of wet nematics, we show the snapshots of nematic order parameter $\mathcal{Q}({\bf r},t)$ (see fig. \ref{fig:S1.1}) and the density field $\rho({\bf r},t) $ (see fig. \ref{fig:S1.2}). In fig \ref{fig:S1.1} (top panel), when $\alpha_1=0.2$ we observe that the $\pm 1/2$ topological defects are distinguishable based on their topological structure. $+1/2$ defects have asymmetric comet like structure and moves along the axis of asymmetry, whereas  $-1/2$ defects have trefoil structure and shows diffusive motion only.  In contrast when $\alpha_1=0.0$, In fig \ref{fig:S1.1} (bottom panel), topological defects are indistinguishable, i.e., they are point like defects. Further, we don't see density growth for $\alpha_1=0.0$ as can be seen when $\alpha_1=0.2$ (see fig. \ref{fig:S1.2}). These structural differences  leads to different defect annihilation mechanisms in  passive and active wet nematics. We again show the correlation length $L_{\mathcal{Q},\rho}(t)$ plots for active and passive cases in fig. \ref{fig:S1.3}. We see that the nematics order parameter follows almost same  growth law for both passive and active wet nematics (see fig. \ref{fig:S1.3}) (left). In contrast, there is  no growth in the density field for the passive case (see fig. \ref{fig:S1.3} (right)).

\section{ Correlation lengths for $\alpha_1=0.3$ and Scaling properties}
\label{appendixB}
\begin{figure}      
{\includegraphics[width=\linewidth]{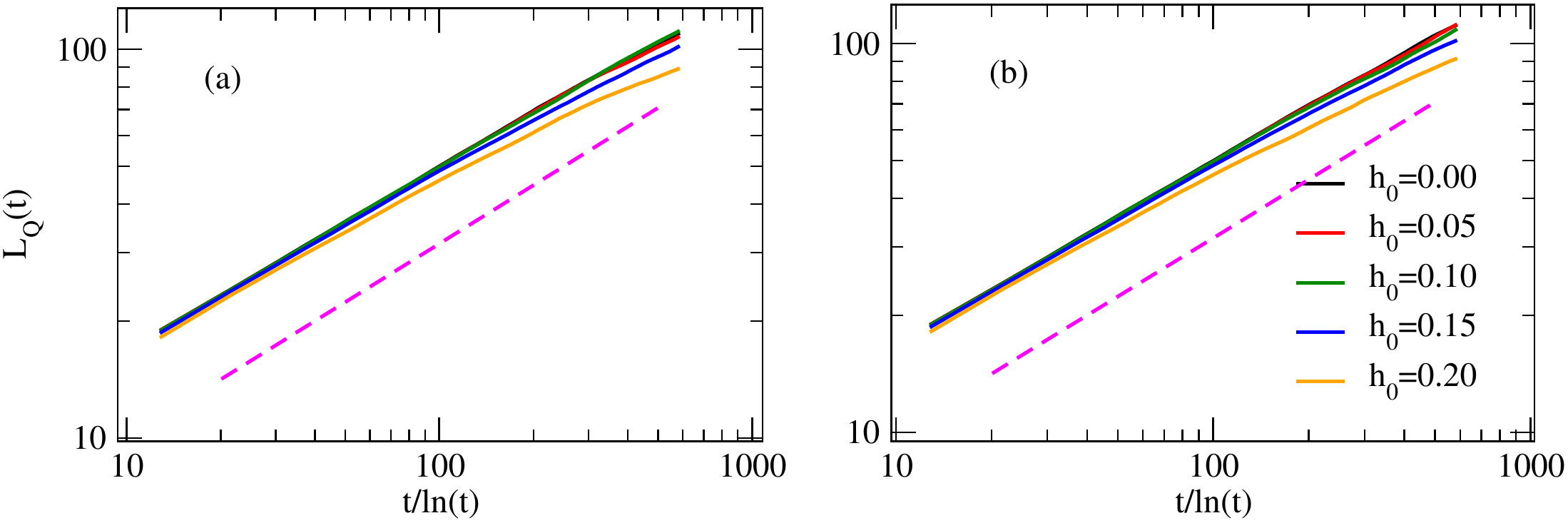}} 
\caption{$L_{\mathcal{Q}}(t)$ vs. $t$   for  $\alpha_1=0.3$ (a) and $\alpha_1=-0.3$ (b) for different values of $h_0$.  }
\label{fig:S1.4}
\end{figure}

\begin{figure}      
{\includegraphics[width=\linewidth]{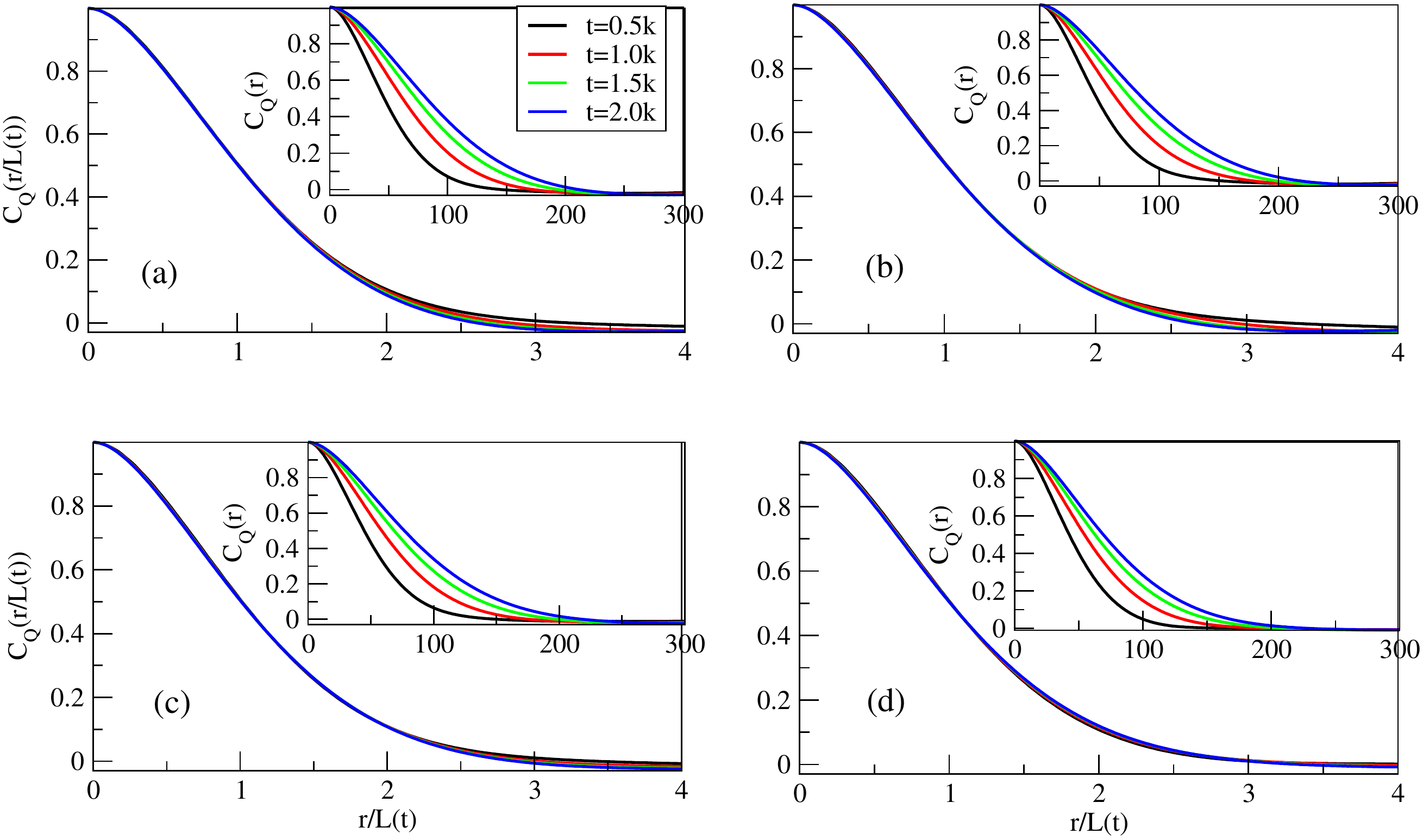}}
\caption{ Scaled two point correlation function, $C_{\mathcal{Q}}({\bf r}/L_{\mathcal{Q}}(t),t)$ vs. scaled distance ${\bf r}/L_{\mathcal{Q}}(t)$  for  $h_0=0.0$ (a), $h_0=0.1$ (b), $h_0=0.15$ (c) and $h_0=0.2$ $\alpha_1=0.2$. Data shown here is for up to time=2000. }
\label{fig:S1.5}
\end{figure}

\begin{figure}      
{\includegraphics[width=\linewidth]{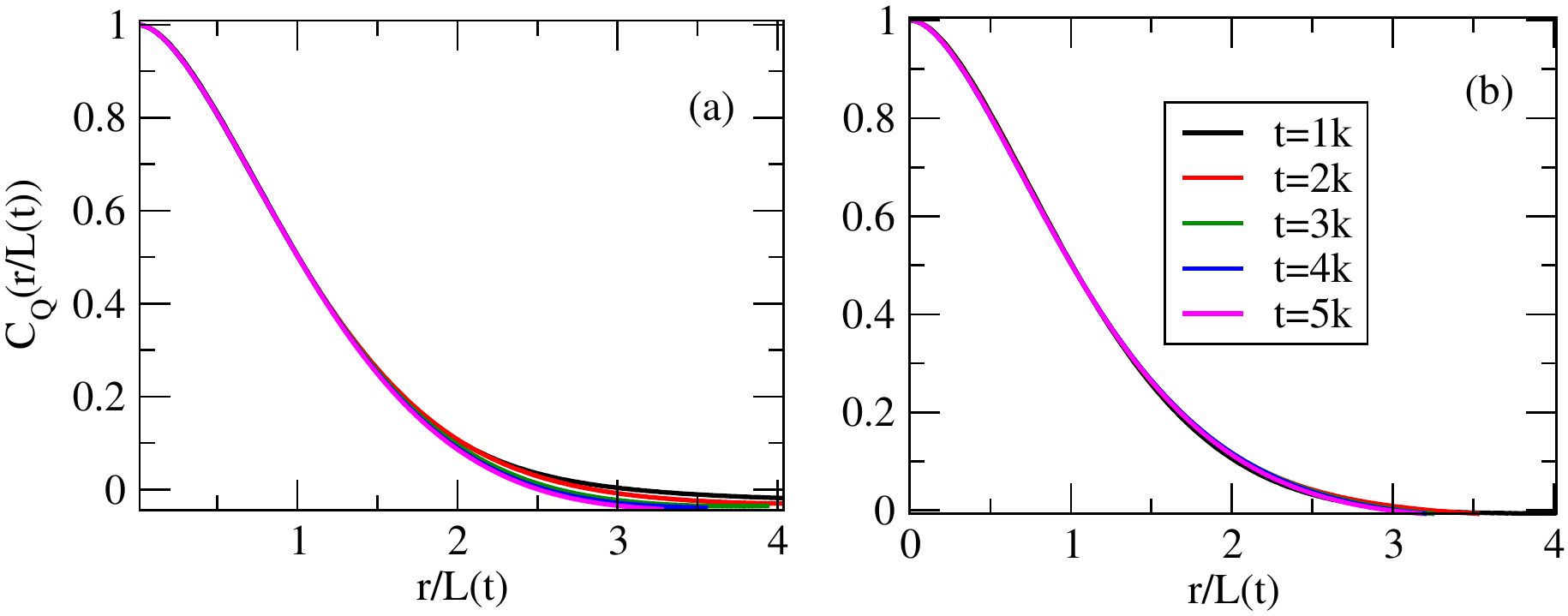}}   
\caption{Scaled two point correlation function, $C_{\mathcal{Q}}({\bf r}/L_{\mathcal{Q}}(t),t)$ vs. scaled distance ${\bf r}/L_{\mathcal{Q}}(t)$  for  $h_0=0.0$ (a) and $h_0=0.2$ (a) for $\alpha_1=0.3$.  }
\label{fig:S1.6}
\end{figure}

\textcolor{black}{In fig \ref{fig:S1.4}, we show the plot of correlation length $L_{\mathcal{Q}}({\bf r},t)$ vs. time $t$ for activity $\alpha_1=0.3$. From this plot, we observe that the observation drawn from fig. \ref{fig:1}, which shows that in RFWAN, disorders response is similar for both contractile and extensile nature of the active stress in the system; is also valid for higher activity in the system.}
\\
fig. \ref{fig:S1.5} show the plot of the early time scaling properties which again confirms that the dynamic scaling improves  as we increase the strength of quenched disorder in the system. Further, fig. \ref{fig:S1.6} shows that this behaviour is also consistent for higher activity in the system.
\\

\section{Numerical Stability check}
\label{appendixC}
\begin{figure}      
{\includegraphics[width=\linewidth]{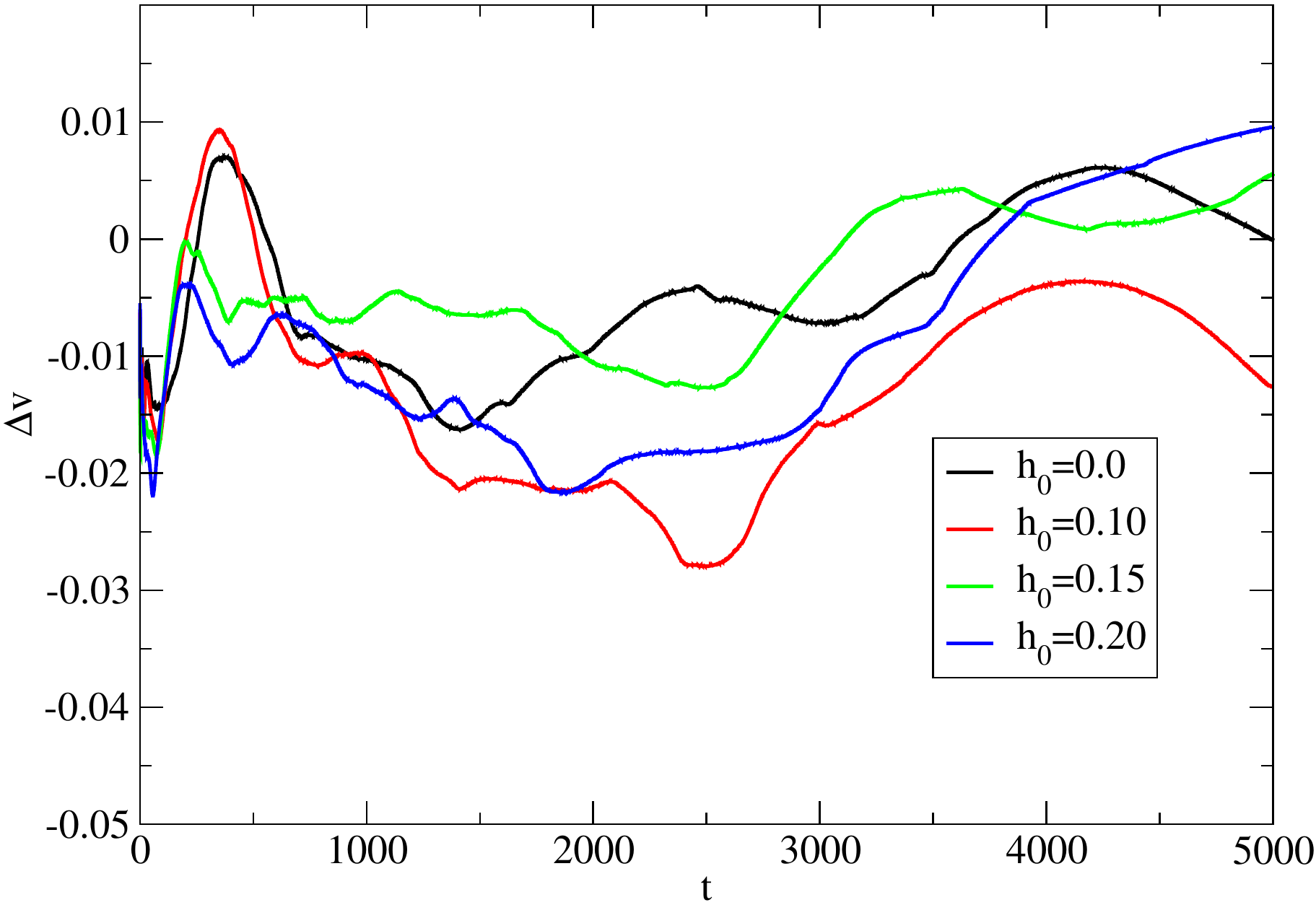}} 
\caption{Fluctuation in velocity field ($v({\bf r},t)$), $\Delta v$ vs. $t$   for different strengths of quenched disorder,  $h_0$.  }
\label{fig:S1.7}
\end{figure}

\textcolor{black}{We check the stability of the code for the chosen set of parameters by calculating the fluctuation in the velocity field $v({\bf r},t)$ from its mean value, i.e., $\Delta v=\langle v \rangle - v_{0}$, where $\langle v \rangle= \sqrt{ \langle v_x^2+v_y^2 \rangle_r}$ and $v_0$ is the mean value of velocity field. In fig. \ref{fig:S1.7}, we plot $\Delta v$ vs. $t$ for different strengths of quenched disorder $h_0$, and observe that $\Delta v$ show small fluctuation as $t \rightarrow \infty$. This implies that the system is stable for the chosen set of parameters.}

\end{document}